\documentclass[10pt,twocolumn]{article}
\usepackage{amsmath}
\usepackage{amsfonts}
\usepackage{amssymb}
\usepackage{graphicx}
\usepackage{bm}
\title{\textbf{Dissociation dynamics in the dissociative electron attachment to ammonia molecule}}
\author{\textbf{Dipayan Chakraborty$^{1\dag}$}, \textbf{Aranya Giri$^{2*}$} and \textbf{Dhananjay Nandi$^{3\dag}$}\\ $^\dag$Indian Institute of Science Education and Research Kolkata, Mohanpur 741246, India\\$^*$ National Institute of Science Education and Research Bhubaneswar,HBNI, Jatni, 752050, India\\ \small{email:$^1$physics.dipayan@gmail.com, $^2$aranya.giri@niser.ac.in, $^3$dhananjay@iiserkol.ac.in}}

\date{}
\begin{document}
\twocolumn[
  \begin{@twocolumnfalse}
    \maketitle
    \begin{abstract}
    Complete dissociation dynamics of low energy electron attachment to ammonia molecule has been studied using velocity slice imaging (VSI) spectrometer. One low energy resonant peak around 5.5 eV and a broad resonance around 10.5 eV incident electron energy has been observed. The resonant states mainly dissociate via H$^-$ and NH$_2^-$ fragments, though for the upper resonant state, signature of NH$^-$ fragments are also predicted due to three body dissociation process. Kinetic energy and angular distributions of the NH$_2^-$ fragment anions are measured simultaneously using VSI technique. Based on our experimental observations, we find the signature of A$_1$ symmetry in the 10.5 eV resonance energy whereas, the 5.5 eV resonance is associated with the well known A$_1$ symmetry.
    \end{abstract}
  \end{@twocolumnfalse}
]
\section{Introduction}
Inelastic electron-molecule collisions lead to production of ions and neutral fragments. Dissociative electron attachment (DEA) is a process in which low energy electron is resonantly captured by the molecule and a temporary negative ion state (TNI) is formed. Subsequently, the resonance state decays into anion and neutral fragment(s). DEA is a topic of interest these days and has been studied by different groups for several molecules \cite{CO:pamir,CO2:pamir,EK_PRL,dea:illenberger}. For example, interaction of high energy radiation with DNA produces low energy secondary electrons, which causes damage to living cells (like single-double strands breaks, DNA-protein cross-links, Mutation, Apoptosis, etc) via DEA to DNA and to its surrounding molecules \cite{dna1, dna2,dea_component, dna_crosslink}. This has provided impetus to rigorous studies on DEA to DNA and respective biomolecules \cite{dea_dna,dea_base,dea_sugar,dea_rna}.
\begin{figure*}
\centering
\includegraphics[scale=.3]{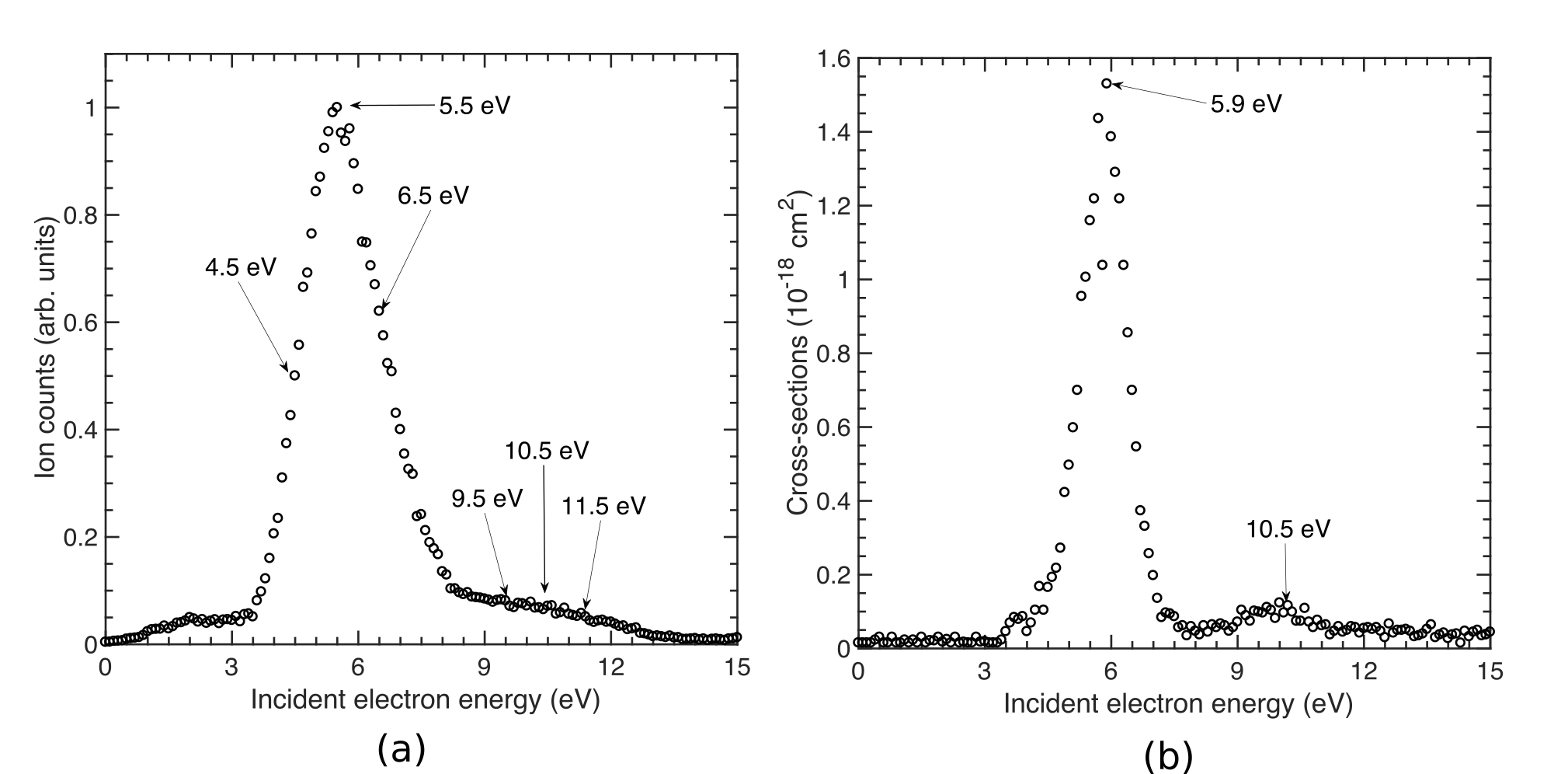}
\caption{Excitation functions of NH$_2^-$ ions obtained from DEA to ammonia. The former one is obtained by using the VSI spectrometer whereas, the second one is from the cross-section spectrometer.}\label{ion yield}
\end{figure*}
Ammonia (NH$_3$) is certainly an essential component for many biological and chemical processes. It is the source of nitrogen for plants (a part of nitrogen cycle), therefore approximately $83\,\%$ of industrial ammonia is used for the production of fertilizers, and it also serves as a raw material for making explosives and cleaning fluids. At the cellular level, its ions are present in nucleic acids. The toxic effect of ammonia can be seen in all animals where it causes neurological dysfunctions \cite{medical}. In Interstellar medium, it is found in the dense molecular clouds of a galaxy \cite{interstellar1} and in grain surfaces \cite{interstellar3,interstellar4}. It is one of the simplest molecules considered while simulating for the production of amino acids in interstellar ice which also gives answers to the generation of life on earth \cite{interstellar2}. 

The study of DEA to ammonia dates back to 1969 when Sharp and Dowell \cite{sharp} and Compton \emph{et al.} \cite{compton} confirmed the two resonances for DEA to ammonia, both producing H$^-$ and NH$^-_2$ ions. At the higher resonance, the presence of the NH$^-$ ion with comparatively lower cross-section was also observed. The cross-sections measured by these two groups differ by a factor of 2 for a particular resonant state anions. Later Rawat \emph{et al.} \cite{rawat} reconfirmed that the resonant states occurred at 5.5 eV and 10.5 eV incident electron energy. The authors also measured the absolute cross-sections for both the resonance state anions using relative flow technique and the reported values for NH$_2^-$ ion for both the resonances are 1.6 $\times$ 10$^{-18}$ cm$^2$ and 0.09 $\times$ 10$^{-18}$ cm$^2$ respectively.

In 1986, Burrow \emph{et al.} \cite{burrow} analyzed the measurements with group theory and suggested that the planar and non-planar dissociation of the lower resonant state results in H$^-$ and NH$^-_2$ decay channels respectively. They also predicted the umbrella mode of oscillation during dissociation. Later, Ram \emph{et al.} \cite{ram} used VSI technique to study the angular distribution (AD) and kinetic energy (KE) distribution of fragment anions. The variation in AD of H$^-$ ions showed that electron attachment preferred particular orientation of ammonia molecule with C$_{3}$ axis along the electron beam direction (from N side to H side). The presence of an umbrella mode of oscillation in lower resonant state was confirmed by the observed variation in AD of the fragment ions with kinetic energy. Using thermochemical and photo-ionization values, the  authors predicted several dissociation channels with their respective threshold energies. Using the AD measurements the authors obtained the symmetry of the TNI state involved in the process as A$_1$ and E for the lower and higher resonance respectively. Recently, Rescigno \emph{et al.} \cite{rescigno} theoretically predicted that at lower resonance, NH$_2^-$ ion is produced by the same H$^-$ ion dissociation channel but through an intermediate \emph{virtual state} where charge is transferred adiabatically at a large internuclear distance. But, in upper resonance there is no mechanism found which is responsible for the formation of NH$_2^-$ ion. In present case, the VSI images have been taken around those resonances to measure the kinetic energy and AD of the fragment ions. From AD measurements, symmetry of the two resonant states is determined.
\section{Instrumentation}
\begin{figure*}
\centering
\includegraphics[scale=.25]{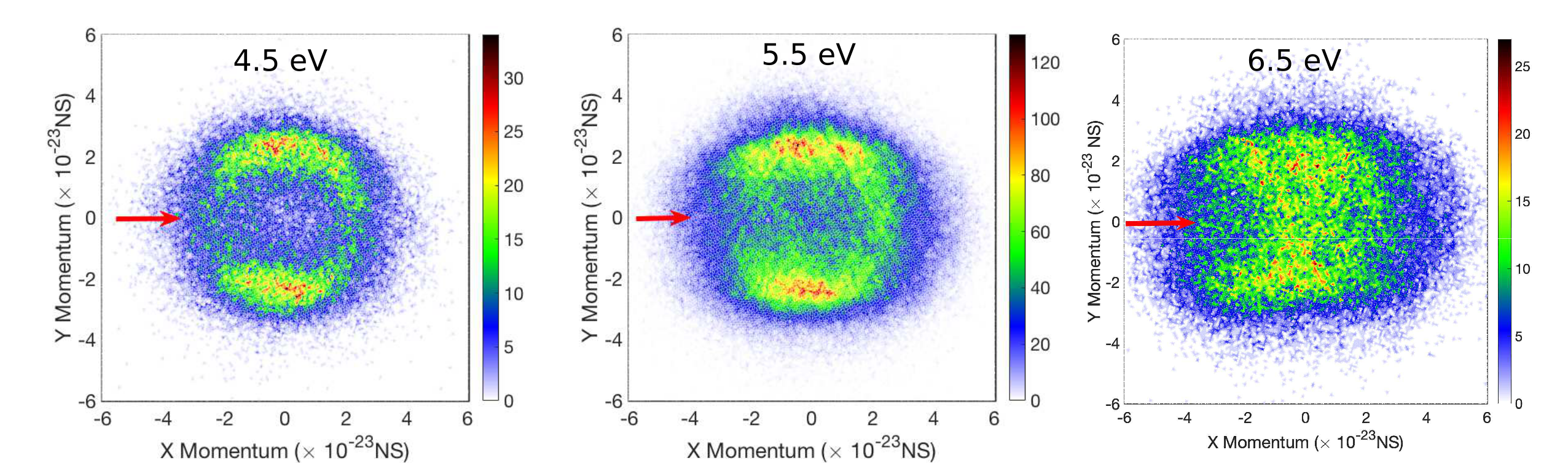}
\caption{Time sliced images taken with a 25 ns time window for NH$_2^-$ formed due to DEA to ammonia molecules for three different electron energy around lower resonance. Electron beam axis is from left to right shown by a red arrow. X-Y direction corresponds to respective momentum in that direction.}\label{lower slice}
\end{figure*}
Details of the experimental setup used for velocity slice images (VSI) are present in different papers \cite{CO2:pamir,dipayan:co}. In the present context we will discuss it briefly. A magnetically collimated pulsed electron beam of 10 kHz repetition rate is produced via thermionic emission process from a tungsten filament with energy resolution around 0.8 eV. This electron beam is made to interact perpendicularly with an effusive molecular beam. When low energy electrons collide with the molecules, negative ion Newton spheres are formed. These Newton spheres are projected to the micro channel plate (MCP) based two dimensional position sensitive detector (PSD) \cite{hex1} by applying moderate extraction field (2 V/cm). The spectrometer is designed to maintain the velocity map imaging (VMI) condition i.e. all the ions formed in the interaction region with a given velocity will map on to a single point on the detector. AD and KE distribution of the fragment ions are obtained from the projections of the Newton spheres. The extraction pulse duration is 2 $\mu$s and is applied 100 ns after the electron beam pulse. This delayed extraction provides sufficient time to expand the Newton sphere so that better time sliced images are extracted. MCP is used to detect the time-of-flight (TOF) of the fragment ions and PSD records the corresponding x and y positions. These x and y positions give momentum information along that direction and the TOF gives the z-momentum. Using CoboldPC software one can record the x and y positions with corresponding TOF for off-line analysis.

To obtain the ion-yield curve, the MCP signal is amplified by a fast amplifier (FAMP) then fed to a constant fraction discriminator (CFD). This CFD signal provides the \textit{stop} signal as input to the time-to-amplitude converter (TAC). The \textit{start} signal is provided by the pulse generator and is synchronized with the electron gun pulse. Time difference between the \textit{start} and \textit{stop} signal determines the TOF of the fragments. Output of the TAC is connected to a multichannel analyzer (MCA). Number of ions hitting the detector is measured by using MCA. Output of the hexanode signals again passes through the FAMP and CFD before it is collected by a time-to-digital converter (TDC) which is directly connected to a computer. Details of this data acquisition system are present in a different paper \cite{MST:pamir}. The aim is to find the central slice of the Newton Sphere, as it contains the kinetic energy and AD information. During off-line analysis, suitable time window is used to select the central one. In the present case 25 ns time window is used for the lower resonance whereas, 50 ns is used for the higher one. Calibration for the kinetic energy distribution measurements has been performed using the kinetic energy released by O$^-/$O$_2$ at 6.5 eV \cite{O2:dhananjay}. Further this energy calibration has been checked by measuring the kinetic energy of the O$^-$ ion produced by DEA to CO$_2$ \cite{CO2:pamir,CO2:slaughter} at 8.2 eV.
\section{Angular distribution of C$_{3v}$ point group}
\label{AD_c3v}
\begin{figure}
\centering
\includegraphics[scale=.3]{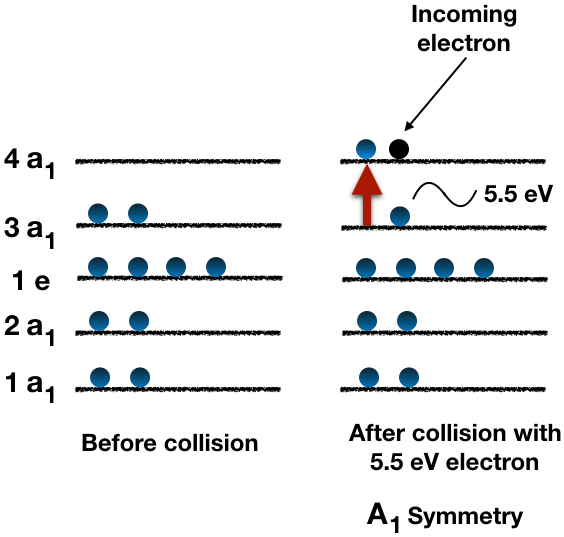}
\caption{Schematic to represent the Feshbach resonance occurred in the NH$_3$ molecule around 5.5 eV. The blue shaded circles represent the electrons present in the parent molecular state, and the black circle represents the incoming electron. Here the incoming electron loses its kinetic energy to excites the 3a$_1$ electron, and both are captured in the 4a$_1$ state. Hence the symmetry of the resonance state is A$_1$.}\label{feshbach_lower}
\end{figure}
The angular distribution of the DEA process is directly related to the symmetry of the TNI state. The dependence of the DEA cross-section of a diatomic molecule as a function of dissociating angle is nicely described by O'Malley and Taylor \cite{malley}. Based on their work, Azria \emph{et al.} \cite{azria} expanded the expression for polyatomic molecules.

Ammonia belongs to C$_{3v}$ point group symmetry. A C$_{3v}$ point group has six symmetry operations, identity (E), rotation of 60$^\circ$ with respect to C$_3$ axis (C$_3$), rotation of 120$^\circ$ with respect to C$_3$ axis (C$_3^2$) and three reflections about three mirror planes formed by the three NH bonds and the C$_3$ axis. Based on the similarities of the operations, there are three symmetry states associated with the C$_{3v}$ point group. The symmetries are A$_1$, A$_2$ and E. Here  A$_1$ and A$_2$ are one dimensional representation whereas, E is two dimensional representation. The ground state configuration of NH$_3$ molecule is A$_1$ so, the transition amplitude from A$_1$ to A$_1$, A$_2$ and E final state transition is calculated by considering various partial waves as,
\begin{equation}
\text{A}_l=\left<\text{Resonant state}|\text{Partial wave}|\text{Initial state}\right>.
\label{transition_amonia}
\end{equation}
Here the partial wave denotes the different partial waves of the incoming electron involved in the transition. This transition amplitude squared and integrated over the azimuthal angle to obtain the variation of DEA cross-section with scattering angle
\begin{equation}
\text{I}(\theta) = \frac{1}{2 \pi} \int_0^{2 \pi} |\text{A}_l|^2 \text{d} \phi 
\label{inti}
\end{equation}
Dissociation occurs in the molecular frame whereas, the measurements of the angular distribution is carried out in the lab frame. So, molecular frame to lab frame transformation of the partial waves for both the incident electron beam and the electronic states is done by the Euler angles ($\phi$, $\theta$, 0) and (0, $\beta$, 0) respectively. Here $\beta$ is the angle between the NH bond and the C$_3$ axis of NH$_3$ molecule. In general, the value of $\beta$ is 68.2$^\circ$, which is used in the present calculations. The character table for the C$_{3v}$ point group along with the symmetry states with corresponding basis functions (described by spherical harmonics) are shown in Table \ref{c3v}. For example, the ground state symmetry of ammonia molecule given by A$_1$ can be expressed by the basis function Y$_0^0$. One can also incorporate more than one partial wave by introducing the phase factor between them. The partial wave approximation used here assumes that the axial recoil approximation is valid, i.e., the dissociation takes place on a time scale before the molecule could undergo rotation or structural changes. The expression for the A$_1$ to A$_1$ final state transitions with three lowest partial waves (s+p+d) and A$_1$ to E final state transitions with two lowest partial waves are given below \cite{ram:thesis}
\begin{table*}[h!]
    \caption{Character Table for C$_{3v}$ symmetry group and their respective basis function}
    \begin{center}
\begin{tabular}{ c c c c c c c c c} \hline
 & & I & & 2C$_3$ & & 3$\sigma_v$ & & Basis Function \\ \hline
A$_1$ & & 1 & & 1 & & 1 & & Y$_{l,0}$ ; l = 0,1,2... \\ 
A$_2$& & 1 & & 1 & &-1 & & $\text{Y}_{3,3} + \text{Y}_{3,-3}$ \\ 
E & & 2 & &-1 & & 0 & & ($\text{Y}_{l,-1}, -\text{Y}_{l,1}$); l=1,2,3...\\ \hline
\end{tabular}
\end{center}
    \label{c3v}
\end{table*}
\begin{align}
\text{I}^{\text{A}_1}_{\text{s+p+d}}(\theta) = & a_0^2 + a_1^2 \big(\sin^2 \beta \sin^2 \theta + 2 \cos^2 \beta \cos^2 \theta \big) \notag \\ & + a_2^2\frac{9}{16} \big(\sin^4 \beta \sin^4 \theta + \sin^2 2\beta \sin^2 2\theta \big) \notag \\ & + a_2^2\frac{1}{8} \big( 3 \cos^2 \beta - 1 \big)^2 \big(3 \cos^2 - 1 \big)^2 \notag \\ & + 4a_0a_1 \cos \beta \cos \theta \cos \delta_1 \notag \\ & + 2a_1a_2\frac{3}{4} \sin \beta \sin 2\beta \sin \theta \sin 2\theta + a_1a_2 \cos \beta \notag \\ & \big( 3 \cos^2 \beta - 1\big) \cos \theta \big( 3 \cos^2 \theta - 1\big)\cos \delta_2 \notag \\ & + a_0a_2\big( 3 \cos^2 \beta -1 \big)\big(3 \cos^2 - 1\big) \cos(\delta_1 + \delta_2)
\end{align}
\begin{align}
\text{I}^{\text{E}}_{\text{p+d}} = & 2 b_0^2 \big( \cos^2 \beta \sin^2 \theta + 2\sin^2 \beta \cos \theta\big) \notag \\  & + \frac{3}{2}b_1^2\big( \frac{1}{4}\sin^2 2 \beta \sin^4 \theta + \cos^2 2\beta \sin^2 2\theta \big) \notag \\ & + \frac{3}{4}b_1^2\sin^2 2 \beta ( 3 \cos^2 \theta - 1 )^2 \notag \\ & + 2b_0b_1\sqrt{3}\cos \beta \cos 2\beta \sin \theta \sin 2\theta \cos \delta_3 \notag \\ & + 2b_0b_1\sqrt{3} \sin \beta \sin 2\beta \cos \theta ( 3 \cos^2 \theta - 1)\cos \delta_3 
\label{A1_E}
\end{align}
All the AD data present in this report are fitted using these two equations.
\section{Results and discussion}
\begin{figure*}
\centering
\includegraphics[scale=.238]{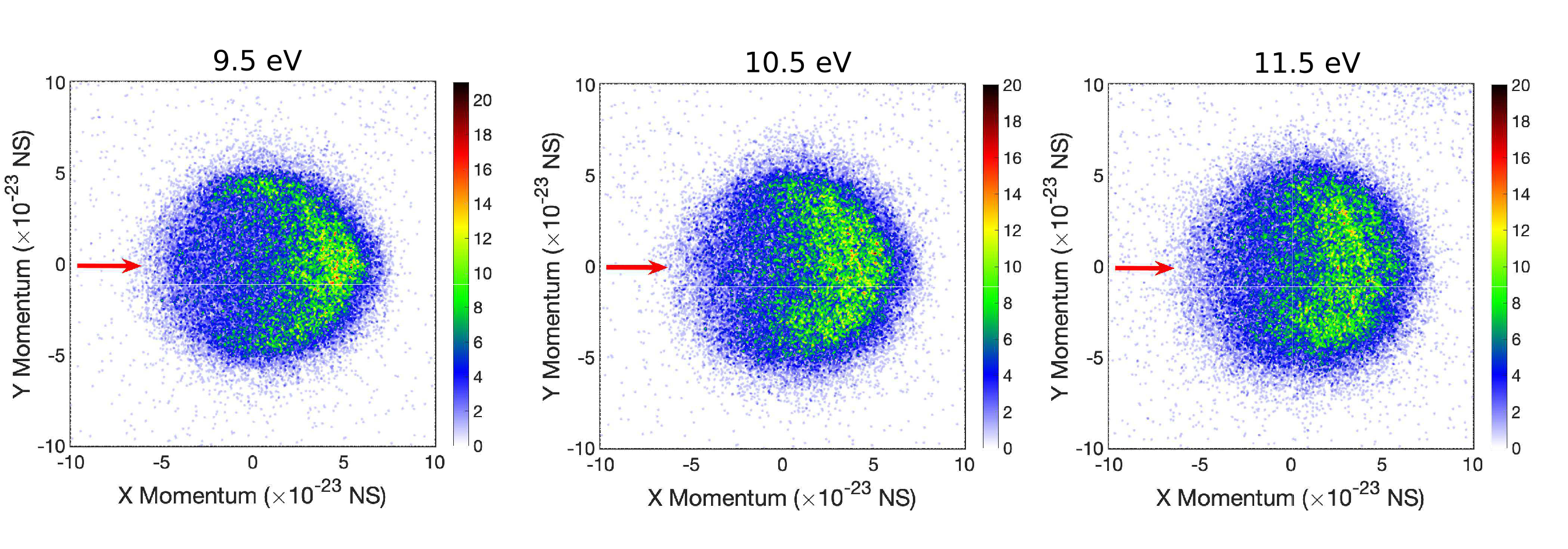}
\caption{Time sliced images taken with a 50 ns time window for NH$_2^-$ ions formed due to DEA to ammonia molecules for three different electron energy around upper resonance state. Electron beam axis is from left to right shown by red arrow. X-Y direction corresponds to respective momentum in that direction.}\label{higher slice}
\end{figure*}

When low energy electron collides with the NH$_3$ molecule, then it resonantly captured by the molecule and forms one temporary negative ion state (TNI), which dissociates via three possible dissociation channels, forming three different negative ions H$^-$, NH$^-$ and NH$_2^-$ :
\begin{equation}
\text{NH$_3$} + \text{e}^- \rightarrow (\text{NH$_3$}^-)^*  \rightarrow \left\{ 
\begin{array}{c}
\text{H}^- + \text{NH$_2$}\\
\text{NH}^- + \text{H} + \text{H}\\
                  \text{NH$_2$}^- + \text{H}
\end{array}
\right.
\end{equation}
\begin{figure}
\centering
\includegraphics[scale=.3]{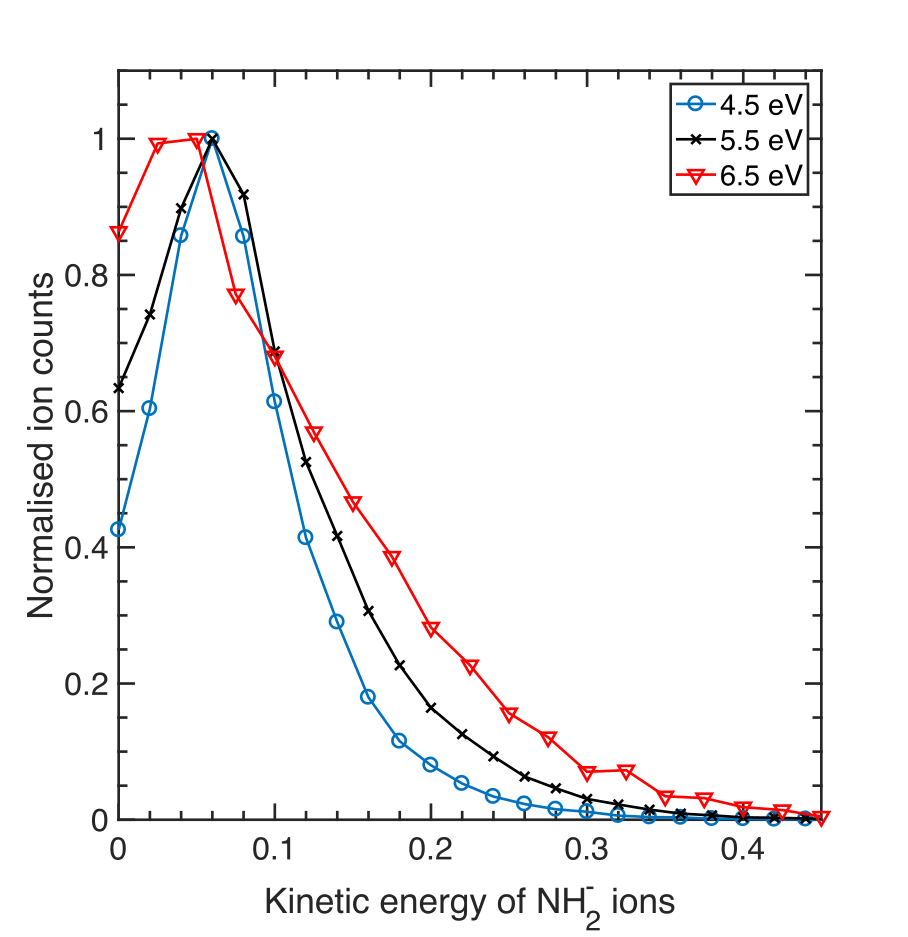}
\caption{The unweighted kinetic energy distribution of NH$_2^-$ ions around lower resonance state for three different electron energies represented by different color.}\label{lower KE}
\end{figure}
Excitation functions of NH$_2^-$ ions obtained from DEA to NH$_3$ are shown in Fig. \ref{ion yield}. The ion yield curves are measured using two different spectrometers. First one is obtained from the VSI spectrometer whereas, the second one is obtained using the cross-section spectrometer \cite{rsi:dipayan}. In the first ion yield curve, the two resonant peaks are overlapped. The possible reason behind this observation could be the poor mass resolution of the VSI spectrometer. In higher resonance, two different fragment anions (NH$^-$ and NH$_2^-$), are resulting into two broad resonant peaks around 10 eV and 10.5 eV respectively \cite{rescigno}. In the second ion yield curve, one can observe two distinct resonant peaks which confirm the presence of temporary negative ion (TNI) states around those energies. To know the kinetic energy and AD of the fragment ions, VSI images are recorded at six different incident electron energies around the two resonances (Fig. \ref{lower slice} and \ref{higher slice}). The incident electron beam axis is from left to right in each image as indicated by a red arrow. For 4.5 and 5.5 eV images (Fig. \ref{lower slice}), one can observe two different lobes perpendicular to the electron beam direction whereas, for 6.5 eV energy the two lobes are not prominent. This indicates the possibility of a different dissociation mechanism involved at this energy. In the previous studies by Rescigno \emph{et al.} \cite{rescigno} and Ram and Krishnakumar \cite{ram}, there is a discrepancy in the 5.5 eV image. Our observation at 5.5 eV energy agreed with the Rescigno's measurement. From the higher resonance images, one can see that the NH$_2^-$ ions are formed mostly in the forward direction of the incident electron beam. A similar observation was made by Rescigno \emph{et al.} where the authors found the H$^-$ momentum distribution and NH$_2^-$ momentum distribution to be exact mirror images. From this observation, the authors concluded that both the fragments are produced from the same resonant state where the negative charge is transferred from the H$^-$ anion to the NH$_2$ fragment at a large internuclear distance. The two resonances, their corresponding dissociation channels, the kinetic energy of the fragment ions and the possible symmetry of the TNI states are discussed below.
\begin{figure}
\centering
\includegraphics[scale=.322]{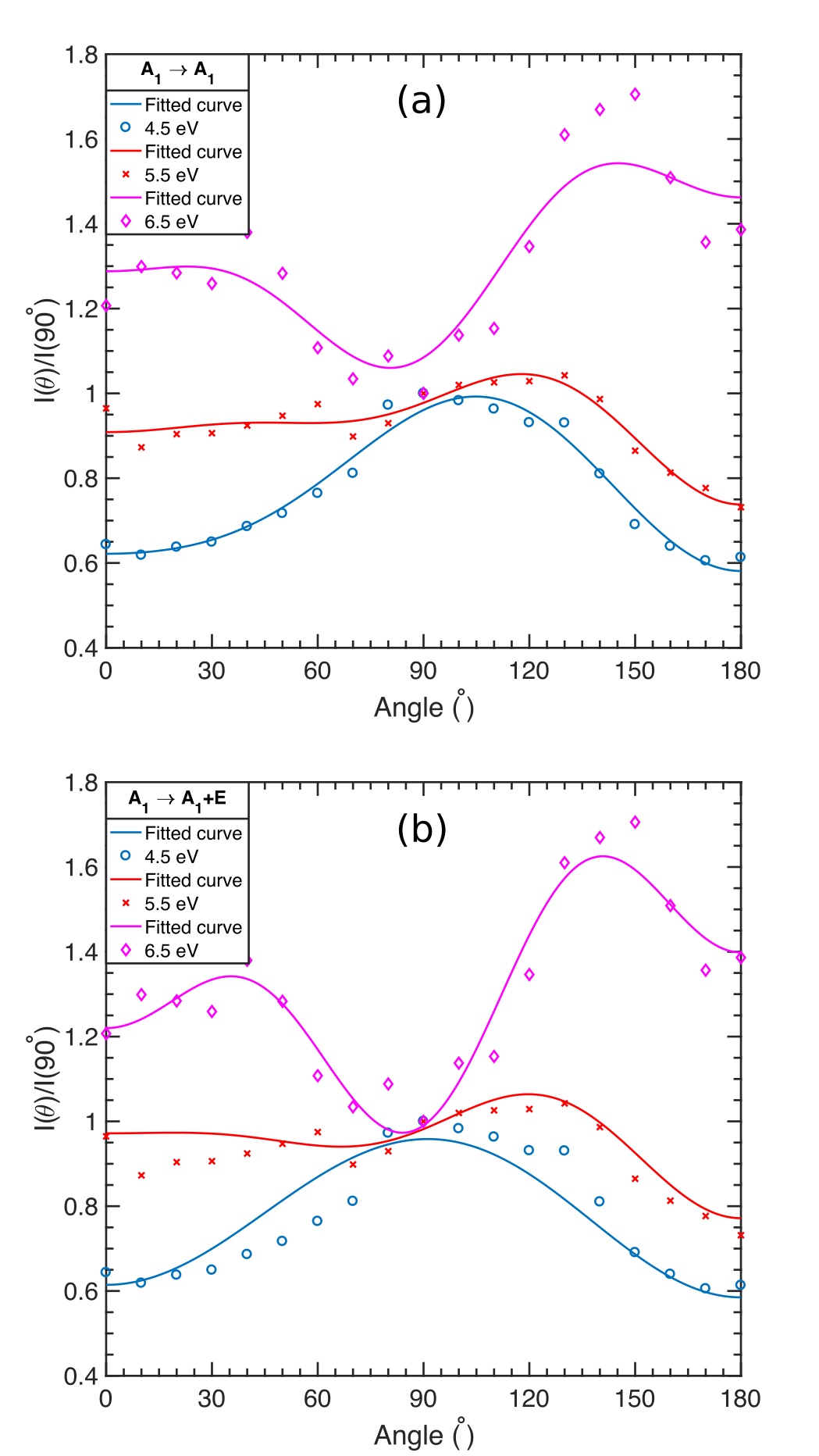}
\caption{Angular distribution of NH$_2^-$ ions (angle with reference to direction of electron beam axis) fitted with (a) A$_1$ to A$_1$ transition, taking s, p, d partial waves, (b) A$_1$ to A$_1 +$E transition, taking s, p, d and p, d partial waves for A$_1$ and E states respectively.}\label{lower AD}
\end{figure}
\subsection{Lower resonance at 5.5 eV incident electron energy}
\begin{table*}[h!]
\caption{Fitting parameters for the angular distribution of NH$_2^-$ ion at lower resonance A$_1$ $\longrightarrow$ A$_1$ transition.}
\begin{center}
     \begin{tabular}{c c c c} \hline
    		& 4.5 eV & 5.5 eV & 6.5 \\ \hline
        Weighting ratio of & & & \\ 
        different partial waves & & & \\
        a$_0$:a$_1$:a$_2$ & 0.83:1:0.11 & 1:0.1:1.31 & 1:.71:2.98\\
        Phase difference(A$_1$) & & & \\
        $\delta^1_{s-p}$ , $\delta^2_{s-d}$ (rad) & 1.65, 3.0 & 1.55, 3.07 & 2.17, 1.68 \\
         & & & \\
        
        R$^2$ value & 0.96 & 0.91 & 0.82 \\ \hline
    \end{tabular}
\end{center}
    \label{A1 table}
\end{table*}
\begin{table*}[h!]
\caption{Fitting parameters for the angular distribution of NH$_2^-$ ion at 6.5 eV for A$_1$ $\longrightarrow$ A$_1 +$ E transition.}
\begin{center}
     \begin{tabular}{c c c c} \hline
        Weighting ratio & Phase & Phase & R$^2$ \\ 
        of different & difference (A$_1$) & difference (E) & value\\
         partial waves &  &  & \\  
        a$_0$:a$_1$:a$_2$ & $\delta^1_{s-p}$ , $\delta^2_{s-d}$ (rad) & $\delta^1_{p-d}$ (rad)& \\
         :b$_0$:b$_1$  & & & \\ \hline
        0.04:1:0.32 &  &  & \\
        0.47:0.55 & 4.46, 3.44 & 1.95 & 0.91 \\ \hline
    \end{tabular}
\end{center}
    \label{6_5_AE}
\end{table*}
The ground state electronic configuration of ammonia molecule is 1a$_1^2\,$2a$_1^2\,$1e$^4\,$3a$_1^2\,$, resulting A$_1$ symmetry. It is already documented that 5.5 eV resonance is a Feshbach resonance where the incoming electron loses its energy to excite the occupied 3a$_1$ valence electron and simultaneously gets captured along with excited electron in the lowest unoccupied (LUMO) 4a$_1$ orbital (Fig. \ref{feshbach_lower}). The dissociation channels produce both H$^-$ and NH$_2^-$ ions \cite{compton,ram,rescigno} in a ratio of 6:4 as mentioned by Rescigno \emph{et al.} \cite{rescigno}. At present we will focus on NH$_2^-$ ion dissociation channel only.
\subsection*{Kinetic energy distribution}
Kinetic energies of the fragment ions obtained from the VSI images are proportional to its radius. So, in order to find the distribution, one should integrate the ion counts over the entire 2$\pi$ angle and plot it with respect to the energy. Fig. \ref{lower KE} shows the kinetic energy distribution of NH$_2^-$ ions, where a peak around 0.06 eV is observed. The constant kinetic energy peak, which is broad in nature with increasing electron energy, reflects internal excitation of the NH$_2^-$ and H fragments. Though, poor electron gun resolution doesn't allow us to separate different kinetic energy bands. The experimentally obtained kinetic energy values of anions are compared with the thermochemical values derived from the given expression
\begin{equation}
\text{KE}_{\text{NH}_2^-}=\left(1-\frac{m}{\text{M}}\right)\left[\text{V}_e-\left(\text{D}-\text{A}+\text{E}^*\right)\right] \label{KE_eqn}
\end{equation}
Here $m$ is the mass of the NH$_2^-$ fragment, M is the mass of the NH$_3$ molecule, V$_e$ is the incident electron energy, D is the NH$_2$-H bond dissociation energy, A is the electron affinity of NH$_2$ atom and E$^*$ is the internal energy of the H atom. From literature, D=4.60 eV \cite{bond_H}, A= 0.77 eV \cite{nh2_affinity} and if we consider the H neutral fragments formed in the ground state, then the thermodynamic threshold of the dissociation channel is 3.83 eV. This dissociation channel was previously observed by Sharp and Dowell \cite{sharp} :
\begin{equation*}
    \text{NH}_3 (\text{A}_1) + \text{e}^- \longrightarrow \text{NH}_3^{-*} (\text{A}_1) \longrightarrow \text{H} (^2 \text{S}) + \text{NH}_2^- (^1\text{A}_1).
\end{equation*}
From Fig. \ref{lower KE}, the NH$_2^-$ ion kinetic energy peak can be observed at 0.06 eV for 5.5 eV resonance. Thus total kinetic energy release (KER) during the process is (total KER =17 times the kinetic energy of NH$_2^-$ ions) 1.02 eV. This indicates that at this resonance NH$_2^-$ ions are produced through the above mentioned dissociation channel, where the threshold is 3.83 eV.
\subsection*{Angular distribution}\label{low_AD}
Fig. \ref{lower AD} shows the AD of NH$_2^-$ ions, extracted from the VSI images for 4.5, 5.5 and 6.5 eV electron energy. For all the incident electron energies, ions with a kinetic energy range between 0-0.15 eV are considered for the AD measurements. The angle is defined with respect to the incident electron beam direction. With close inspection, one can observe that most of the ions are concentrated within 50$^\circ$ to 150$^\circ$ for all the energies. The VSI images are anisotropic and are found to be same as the LBNL and Heidelberg experimental VSI images. It was believed that in this energy region, separate dissociation channels resulting both H$^-$ and NH$_2^-$ ions are present until Rescigno concluded that NH$_2^-$ ions are formed due to an adiabatic charge transfer from H$^-$ to NH$_2$ at large inter-nuclear distance, i.e., through a virtual-state channel \cite{rescigno}. This makes the AD of NH$_2^-$ as the AD of H$^-$ reflected through $90^o$ from the electron beam axis ($\theta_{\text{NH}_2^-} = 180^\circ - \theta_{\text{H}^-}$). We took the AD of H$^-$ from the measurements of Ram \cite{ram} and Rescigno \cite{rescigno} for 4.5 eV and 5.5 eV as a reference and compared it with our NH$_2^-$ AD curve. At 4.5  and 5.5 eV energies, the AD of H$^-$ ion peaked at $85^o$ and $70^o$ respectively. Thus in the present measurements for 4.5 eV and 5.5 eV incident electron energy $\theta_{\text{NH}_2^-}$ is equal to $95^\circ$ and $120^\circ$, confirming the virtual-state channel. Fig. \ref{lower AD} also shows that when electron energy increases, backward scattering increases, which becomes more dominant at higher energy. The observed broad AD can be explained due to the umbrella mode vibration of the TNI state present in this resonance \cite{burrow,ram}. The AD results can be discussed further with respect to the basic structure of ammonia molecule. It has a pyramidal shape with the N atom situated at the top and three H atoms at the base. The basic symmetry of the molecule is C$_{3v}$ where the C$_3$ axis is passing through the N atom and the center of the triangle formed by the three H atom. The angle between the NH bond and the C$_3$ axis is 68.2$^\circ$ and the 3a$_1$ orbital which is excited during this 5.5 eV resonance has the electron density distributed up and down the N atom. Now during the dissociation of the TNI, preferential direction for the higher energetic H$^-$ ion is along the N-H bond axis. From the AD measurements it is observed that the NH$_2^-$ ions are peaking at 120$^\circ$ direction i.e. the H$^-$ ion is at 70$^\circ$ (close to 68.2$^\circ$). This clearly implies that the preferential orientation of the ammonia molecule during the electron attachment process is along the C$_3$ axis, from N to H direction.

To know the symmetry of the associated TNI state, the AD data is fitted with the theoretical expression as discussed in section \ref{AD_c3v}. Fig. \ref{lower AD} (a) shows the fitted AD curve for A$_1$ to A$_1$ transition. It can be observed that the fitted AD curve is enough to claim that the symmetry of the resonant state involved is A$_1$. Slight deviation is observed for the 6.5 eV energy. To investigate the possible involvement of any other symmetries, data points are fitted with A$_1$ to A$_1 +$E transition model, which provides a better-fitted AD curve for 6.5 eV energy. Thus one can predict involvement of E symmetry state around 6.5 eV energy region. Expression \ref{A1_E} represents A$_1$ to E transition model for the lowest two partial waves \cite{ram:thesis}. The values of different parameters used in the fit function are listed in Table \ref{A1 table} and Table \ref{6_5_AE} with corresponding R$^2$ values.
\begin{figure}
\centering
\includegraphics[scale=.3]{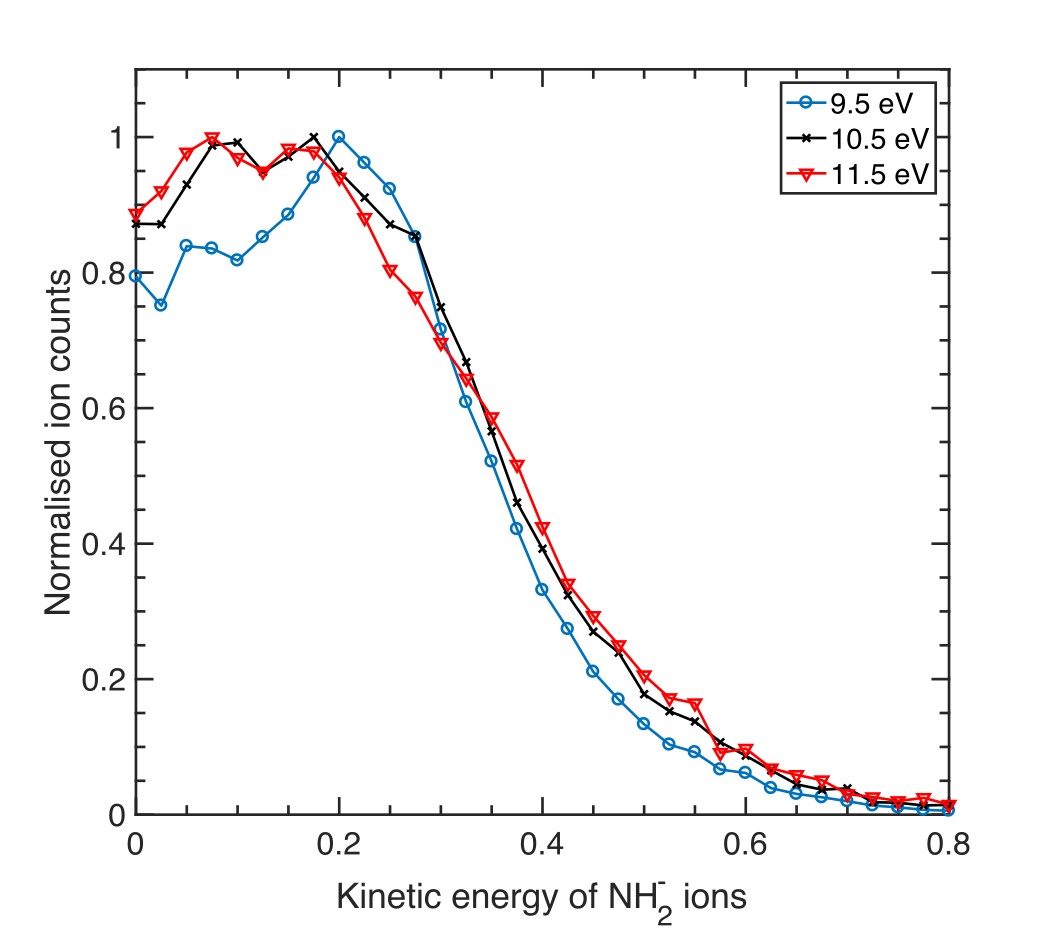}
\caption{The unweighted kinetic energy distribution of NH$_2^-$ ions around the upper resonance state for three different electron energies.}\label{higher KE}
\end{figure}
\subsection{Higher resonance at 10.5 eV incident electron energy}
The dynamics involved in the higher resonance is not as simple as the lower one. To describe the dynamics involved in this resonance process, Ram and Krishnakumar \cite{ram} compare it with the VUV absorption and photo-electron spectrum \cite{walsh,robins} of ammonia molecule where $1e \rightarrow3sa_1$ Rydberg transition occurred at 10.6 eV energy. This result leads them to think that it is a Feshbach resonance where the HOMO-1 valence 1$e$ electron excites and simultaneously two electrons are captured in the LUMO 4a$_1$ orbital. As a result, the symmetry of the resonance state involved in the process is E. But from the NH$_2^-$ ion AD behavior, the authors could not find any robust signature of E symmetry in the resonant state. This contrast between the understanding and the experimental observation is described by the authors. The double degeneracy of the 1e orbital could be manifested as Jahn-Teller effects or other non-adiabatic effects which lead to the rapid distortion of the molecular geometry. As a result, the AD data does not clearly reflect the E symmetry involved in the process. Now, if the distortion of the molecular geometry is the reason, then both H$^-$ and NH$_2^-$ AD shouldn't reflect the E symmetry. But, their H$^-$ AD reflects the involvement of E symmetry at the same resonance energy. Later, in the theoretical study by Rescigno \emph{et al.} \cite{rescigno}, it was found that axial recoil approximation breakdown is less severe in this resonance as there is no barrier to direct dissociation. In their experimental and theoretical study, the authors confirmed that H$^- +$ NH$_2$ dissociation channel occurs due to the involvement of $^2$E symmetry state which is in agreement with the present understanding. But they are unable to locate any dissociation channel resulting to NH$_2^-$ ions if the symmetry of the TNI state is E. Hence they termed the presence of the NH$_2^-$ ions in the upper resonance state as a mystery. So, the dynamics of the upper resonance state in ammonia is still an open question with its symmetry and possible dissociation channels. To address this problem, we took the study of resonance enhanced multi-photon ionization (REMPI) spectrum obtained by Langford \emph{et al.} \cite{langford}. Here the authors found a 1a$_1''\rightarrow$5pa$_2''$ Rydberg transition to occur within the energy range 9.5 to 10.1 eV. The authors represent the states in C$_s$ symmetry and the A$_2$ state in C$_s$ symmetry will be either A$_1$ or E symmetry on making a comparative solution comparing the point group C$_s$ to C$_{3v}$. This Rydberg state can be a parent state for the Feshbach resonance that occurred within this energy region. As a consequence one electron from the valence 3 a$_1$ is excited to a higher a$_1$ orbital and captured along with the incoming electron. As a result, the symmetry of the TNI state will be A$_1$, which can be a parent state for the NH$_2^-$ dissociation channel. Our experimental observation clearly shows the involvement of A$_1$ symmetry in this resonance.
\subsection*{Kinetic energy distribution}
Fig. \ref{higher KE} represents the kinetic energy distribution of the fragment ions formed around 10.5 eV resonance energy. For 9.5, 10.5 and 11.5 eV, one small peak at around 0.2 eV is observed. The broad distribution can be in due to the internal excitation of the NH$_2^-$ ions. Sharp and Dowell \cite{sharp} speculated that NH$_2^-$ ion produced in this energy range is in its first electronically excited state. Though, dissociation channels for ground state NH$_2^-$ ions were also predicted in this resonance energy \cite{ram}. By using the thermochemical values in Equation (\ref{KE_eqn}), the maximum kinetic energy for ground state NH$_2^-$ ion is found to be 0.4 eV. But observed maximum kinetic energy, in this case, is about 0.55 eV, which is within the electron gun energy resolution range. Another possible contribution to kinetic energy distribution is due to the presence of NH$^-$ ions. With respect to this, if NH$^-$ ions are formed with 0.2 eV kinetic energy, total kinetic energy release by the process will be 1.7 eV. This clearly indicates the presence of H + H + NH$^-$($^2\Pi$) channel, whose thermodynamic threshold is 8.08 eV \cite{ram}. Thus from kinetic energy measurements, one can predict the presence of a three-body dissociation channel in this resonance however, due to poor mass resolution capability of the VSI spectrometer we are unable to separate NH$^-$ and NH$_2^-$ fragments.
\begin{figure}
\centering
\includegraphics[scale=.26]{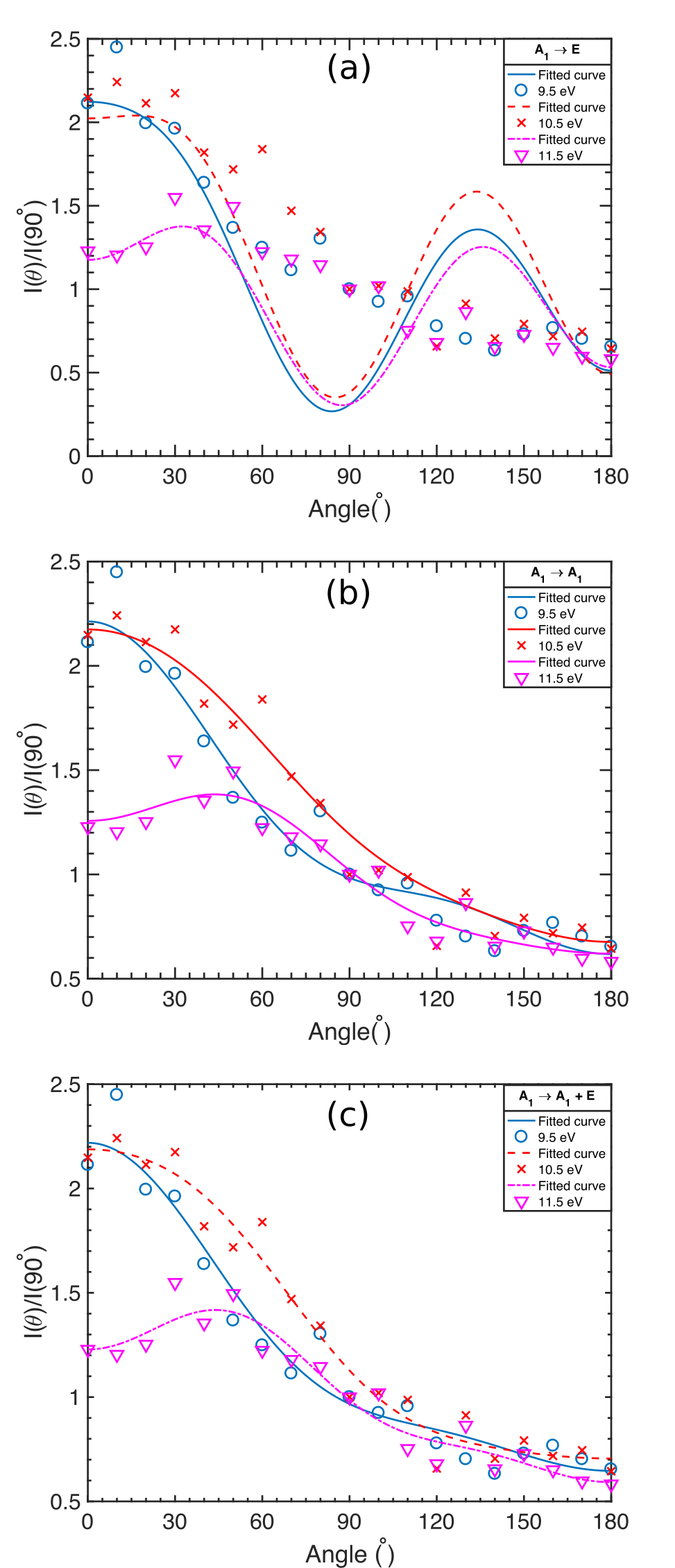}
\caption{Angular distribution of NH$_2^-$ ions (angle with reference to direction of electron beam axis) fitted with (a) A$_1$ to E final state transition model, considering lowest two (p and d) partial waves, (b) A$_1$ to A$_1$ final state transition model with lowest three (s, p, d) partial waves, (c) A$_1$ to A$_1 +$E final state transition model, taking s, p, d and p, d partial waves for A$_1$ and E states respectively.} \label{higher AD}
\end{figure}
\begin{figure*}
\centering
\includegraphics[scale=.226]{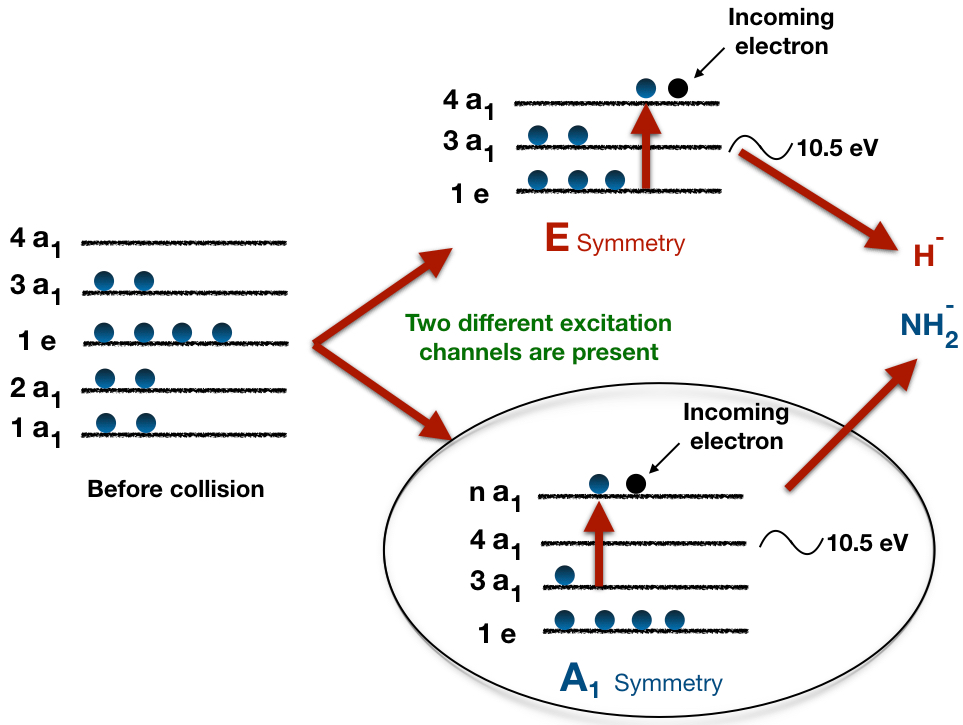}
\caption{Schematic to represent the Feshbach resonance occurred in the NH$_3$ molecule. The blue shaded circles represent the electrons present in the parent molecule, and the black circle represents the incoming electron. At 10.5 eV resonance, two different excitation channels are present. The former one is through the excitation of an 1e electron and subsequently capture of two electrons to the 4a$_1$ orbital, resulting an E symmetry of the resonant state (This channel has been observed previously). Whereas the latter one is through the excitation of an 3a$_1$ electron and capture of two electrons into a higher a$_1$ orbital. Hence, the resulting symmetry of the resonant state is A$_1$. The E symmetry state is responsible for the formation of H$^-$ ions whereas, the A$_1$ symmetry state is responsible for the formation of NH$_2^-$ ions.}\label{feshbach_higher}
\end{figure*}
\subsection*{Angular distribution}
\begin{table*}[h!]
\caption{Fitting parameters for the angular distribution of NH$_2^-$ ion at upper resonance A$_1$ $\longrightarrow$ A$_1$ transition.}
\begin{center}
     \begin{tabular}{c c c c} \hline
    		& 9.5 eV & 10.5 eV & 11.5 \\ \hline
        Weighting ratio of & & & \\ 
        different partial waves & & & \\
        a$_0$:a$_1$:a$_2$ & 2:1:0.84 & 1.36:1:0.97 & 0.70:1:1.36\\
        Phase difference(A$_1$) & 0.34,2.37 & 0.61,1.60 & 0.74,1.45 \\
        $\delta^1_{s-p}$ , $\delta^2_{s-d}$ (rad) & & & \\
         & & & \\
        
        R$^2$ value & 0.95 & 0.97 & 0.93 \\ \hline
    \end{tabular}
\end{center}
    \label{A_A_higher}
\end{table*}
\begin{table*}[h!]
\caption{Fitting parameters for the angular distribution of NH$_2^-$ ion at 9.5 eV for A$_1$ $\longrightarrow$ A$_1 +$ E transition.}
\begin{center}
     \begin{tabular}{c c c c} \hline
        Weighting ratio & Phase & Phase & R$^2$ \\ 
        of different & difference (A$_1$) & difference (E) & value\\
         partial waves &  &  & \\  
        a$_0$:a$_1$:a$_2$ & $\delta^1_{s-p}$ , $\delta^2_{s-d}$ (rad) & $\delta^1_{p-d}$ (rad)& \\
         :b$_0$:b$_1$  & & & \\ \hline
        0.9:1:0.18 &  &  & \\
        :0.6:0.01 & 0.54, 3.08 & 0.03 & 0.98 \\ \hline
    \end{tabular}
\end{center}
    \label{9_5_AE}
\end{table*}
To know the symmetry of the associated resonance state, AD of the NH$_2^-$ ions is extracted from the VSI images. From the AD data it can be observed that for 9.5 and 10.5 eV energies, most of the ions are formed in the forward direction. The AD data is fitted with the procedure as discussed in Section \ref{AD_c3v}. Following the same procedure, we fit our AD data for A$_1 \rightarrow$ A$_1$, A$_1 \rightarrow$ E and A$_1 \rightarrow$ A$_1 +$E final state transition.

The results for the A$_1 \rightarrow$ E and A$_1 \rightarrow$ A$_1$ final state transitions are shown in Fig. (\ref{higher AD}a) and (\ref{higher AD}b). Here the lowest two partial waves for E final state transition and lowest three partial waves for A$_1$ final state transition are considered because the contributions of the higher partial waves are increasingly small. From the fitted AD curve, it is clear that A$_1 \rightarrow$ E transition is inadequate to reflect the symmetry of the TNI state whereas, A$_1 \rightarrow$ A$_1$ final state transition gives us a better fit with good R$^2$ value (over 0.9) which clearly indicates that A$_1$ state is present in this resonance. From the current experimental understanding and from the previous studies we propose that two closely lying resonant states with symmetry A$_1$ and E are present within this 10.5 eV resonance. The H$^-$ ions are formed due to the E symmetry state whereas, the NH$_2^-$ ions contribution came mainly due to the A$_1$ symmetry state. It will be interesting to see whether the A$_1$ symmetry is also responsible for the H$^-$ ions. But with the current experimental facilities, it is not possible to detect the H$^-$ ions. The fitted AD data for A$_1 \rightarrow$A$_1 +$ E final state transition is also shown in Fig. (\ref{higher AD}c), which is almost the same as Fig. (\ref{higher AD}b). Hence the E state contribution can be ruled out. Only for 9.5 eV energy, slightly better-fitted AD curve is observed for A$_1 \rightarrow$ A$_1 +$E transition. This little contribution from E state can be in due to the presence of NH$^-$ ions via three body dissociation process, which is possible in this energy region. The fitting parameters for A$_1 \rightarrow$A$_1$ and A$_1 \rightarrow$A$_1 +$E transition are given in Table \ref{A_A_higher} and \ref{9_5_AE}. Theoretical calculations by Dr. P. C. Minaxi Vinodkumar (private communications) confirms the presence of A$_1$ and/or E state around 10.2 eV energy region which, further supports our conclusion \cite{vinodkumar}. It is to be mentioned here that, the dissociation dynamics of ammonia molecule in higher resonance is complex. When the Rydberg transition occurs, the NH$_3$ molecule no longer holds the C$_{3v}$ symmetry. So the resonance starts with a C$_{3v}$ geometry before it goes to some other symmetry. A high-level time-dependent theoretical calculation is imperative to understand the dynamics properly.
\section{Conclusion}
Complete DEA dynamics of ammonia molecule is studied using VSI technique. Two resonances around 5.5 eV and 10.5 eV are observed. The VSI images of NH$_2^-$ fragment ions are measured around the 5.5 eV and 10.5 eV resonance energies. KE distribution and AD of the NH$_2^-$ ions are extracted from the slice images. From the KE and AD measurements, we reconfirm the presence of A$_1$ symmetry in the 5.5 eV resonance energy. KE distribution of the 10.5 eV resonance indicates the involvement of three body dissociation process. Our AD measurements clearly indicates the presence of A$_1$ symmetry state in the 10.5 eV resonance.
\section{Acknowledgements}
D. N. gratefully acknowledges the partial financial support from ``Science and Engineering Research Board (SERB)'' under the project No. ``EMR/2014/000457''. DC is thankful to IISER Kolkata for providing research fellowship.

\bibliographystyle{h-physrev}
\bibliography{nh3bib.bib}

\end{document}